# A Non-Biased Trust Model for Wireless Mesh Networks

Heng Chuan Tan[1], Maode Ma [1], Houda Labiod[2], Peter Han Joo Chong[3], Jun Zhang [2]

[1]School of Electrical and Electronic Engineering, Nanyang Technological University, Singapore
{htan005, emdma, ehjchong} @ntu.edu.sg

[2]INFRES, Telecom ParisTech, 46, rue Barrault, 75013, Paris cedex 13, France
{Labiod,jun.zhang}@telecom-paristech.fr

[3]Department of Electrical and Electronic Engineering, Auckland University of Technology, Auckland, New Zealand
{peter.chong}@aut.ac.nz

**Abstract**

Trust models that rely on recommendation trusts are vulnerable to badmouthing and ballot-stuffing attacks. To cope with these attacks, existing trust models employ different trust aggregation techniques to process the recommendation trusts and combine them with the direct trust values to form a combined trust value. However, these trust models are biased as recommendation trusts that deviate too much from one's own opinion are discarded. In this paper, we propose a non-biased trust model that considers every recommendation trusts available regardless they are good or bad. Our trust model is based on a combination of two techniques: the dissimilarity test and the Dempster Shafer Theory. The dissimilarity test determines the amount of conflict between two trust records, whereas the Dempster Shafer Theory assigns belief functions based on the results of the dissimilarity test. Numerical results show that our trust model is robust against reputation based attacks when compared to trust aggregation techniques such as the linear opinion pooling, subjective logic model, entropy-based probability model and regression analysis. In addition, our model has been extensively tested using network simulator NS-3 in an Infrastructure based WMN and a Hybrid based WMN to demonstrate that it can mitigate blackhole and grayhole attacks.

# I. INTRODUCTION

A Wireless Mesh Network (WMN) is formed by a collection of mesh routers and mesh clients that cooperate to establish mesh connectivity and network coverage over an area [1, 2]. The mesh routers are usually static and form the wireless mesh backbone of the network for providing network access. The mesh clients, on the other hand, can be static or mobile and have limited computing capabilities. Each node relies on the cooperation of the intermediate nodes in forwarding the packets to the destination along a multi-hop path. The advantages of WMNs include low deployment cost, easy extension of network coverage to hard-to-wire areas and highly reliable wireless connectivity via the multi-hop communications. Due to this, WMN has found widespread applications in many areas of wireless networking that include public safety/military, residential, enterprise, campus networking and in rescue operations. Depending on the application requirements and the type of nodes present, WMNs can be further classified into an Infrastructure based WMN, Client based WMNs and Hybrid based WMNs where the Hybrid based WMNs is basically a combination of the Infrastructure based WMN and Client based WMNs architecture consisting of static mesh routers and mobile clients respectively.

However, due to the openness and the multi-hop nature of the WMNs, attacks can be launched at any layer of the Internet protocol stack [3, 4]. While cryptography has been the traditional approach of providing network security in terms of confidentiality, integrity, authentication, and non-repudiation, it is not sufficient to protect the network against nodes exhibiting selfish behaviors i.e. nodes that refrain from forwarding packets to save scarce resources. Such selfish behavior is hard to detect as they appear to be legitimate nodes with valid cryptographic keys. Subsequently, they may misbehave and conduct packet dropping attacks such as blackhole and grayhole attacks to degrade the overall throughput of a network. To enforce cooperation in the network, various trust models have been developed to detect and isolate selfish nodes in order to improve the network throughput. The first pioneering work [5] was proposed in 2000 by Marti et al., who proposed the use of trust ratings to assess the reliability of a node in forwarding packets. In their model, they have proposed the Watchdog mechanism which gathers information based on the promiscuous mode of observations. Using this information, a trust rating is defined for selfish node detection. The downside of this model is that trust convergence is slow as the trust evaluation only depends on the direct observations.

To enable the trust value to converge faster, [6-8] have proposed to rely on both the direct trust and indirect trusts from the neighboring nodes. By gathering recommendation trusts from the other nodes, it improves the detection time of selfish nodes in the network [9]. Consequently, the use of direct and indirect trusts has become the design standards of many recent trust models. However, as pointed out by [10-14], trust models that depend on recommendation trusts are more prone to badmouthing attacks and ballot-stuffing attacks. More specifically, the malicious recommender may assign a very poor trust value in an attempt to demote the reputation of a well-behaved node known as the badmouthing attacks. As a result, the well-behaved node is blacklisted from further communications which lead to network partition. The malicious recommender may also try to promote its accomplice's trust level by recommending a very high trust value known as the ballot-stuffing attacks. This prolongs the lifetime of the malicious nodes in the network for causing further damage. Methods of preventing the badmouthing and ballot-stuffing attacks have been discussed by [9-22] where models employing different trust aggregation techniques to combine the direct trusts and indirect trusts have been proposed. These techniques impose further constraints to process the recommendation trusts before the aggregation process. In [9], recommendation trusts are only considered if they pass the deviation test which is determined by the similarity with one's own direct trust. In [7, 13], only positive recommendation trusts are allowed to propagate in the network. In [10-12, 14], constraints are imposed where only recommendation trusts from the trustworthy nodes with trust value higher than a pre-defined threshold are considered in the trust aggregation. Due to these reasons, the existing trust models are said to be biased as important evidence may be discarded. In addition, trustworthy nodes may misreport the recommendation trusts only to a certain group of nodes in the network which increase the difficulty of detection.

In this paper, we propose a non-biased trust model called the DS-Trust model to address two problems in WMNs, that is, (1) to overcome the attacks that can subvert the proper operation of the reputation systems, namely the badmouthing and the ballot-stuffing attacks (2) to mitigate the effects of packet dropping attacks that cannot be solved by cryptography. Our trust model is non-biased because every recommendation trusts, whether they are good or bad are considered in the trust aggregation process using the Dempster Shafer Theory (DST) [23]. Henceforth, our trust model is called the DS-Trust model. The major contributions of our work are summarized below:

- Propose the Dempster Shafer's rule of combination to fuse direct trust and indirect trusts together to obtain the final combined trust.

- Introduce a dissimilarity test to check the amount of conflict between one's own observations and each received recommendation trusts. The result of the dissimilarity test is then used to determine the basic probability assignment (bpa) mass for each subset of the power set in DST.

- Demonstrate that DS-Trust has a higher resiliency against badmouthing attacks and ballot-stuffing attacks compared to the benchmarking approaches.

- Incorporate the DS-Trust into the Ad hoc On Demand Distance Vector Routing (AODV) protocol [24] and analyze the performance of an Infrastructure based WMN and Hybrid based WMNs to show that it can mitigate packet dropping attacks.

The rest of the paper is organized as follows. Section II reviews the related works on trust aggregation techniques. Section III presents the network model and reviews some of the attacks addressed by DS-Trust. Section IV presents the concept of DST that serves as a basis of our proposed trust model. Section V provides the details of the DS-Trust model. Section VI presents the performance improvements of DS-Trust against badmouthing and ballot-stuffing attacks. Section VII presents the simulation results to demonstrate the performance of DS-Trust in an Infrastructure based WMN and Hybrid based WMN. Section VIII concludes the paper.

## II. RELATED WORKS

Recommendation trusts are important in trust modeling because it improves the trust evaluations and shortens the detection time of selfish nodes in the network. When recommendation trusts are available, there needs to be a method to combine them with the direct trust values. This process is commonly known as the trust aggregation. In this section, we provide a review of the most common trust aggregation techniques and classify them into four types: linear opinion pooling, entropy-based probability model, subjective logic (SL) operators and regression analysis.

### A. Linear Opinion Pooling

In [9, 16], S. Buchegger and J.-Y. Le Boudec have proposed the use of linear opinion pooling to combine the recommendation trusts to form an indirect trust value. In this approach, the linear opinion pooling technique is just a weighted average of the individual recommendation trusts as shown in (1).

$$T_{i,j}^{Indirect} = \sum_{k=1}^{N} \omega_k T_{k,j}^{direct} \qquad \forall k = 1,2,...,N \qquad (1)$$

where $\omega_k$ is a small positive weight in the range (0,1), $T_{k,j}^{direct}$ denotes the recommendation trust of node $j$ as observed by recommender $k$ and $T_{i,j}^{Indirect}$ denotes the indirect trust of node $j$ as derived by node $i$. In this model, the recommendation trusts are only considered if they are compatible with the current indirect trust value determined using a deviation test, otherwise the recommendation trusts are discarded. In [13, 20, 21], the authors have proposed to derive the weight associated with each recommendation trust in (1) based on the trustworthiness of the recommender such that the recommendation trusts that come from an untrustworthy recommender are discounted more than the one that comes from a highly reputed recommender. In [21], Li et al. have proposed an additional check besides the deviation test to differentiate badmouthing and conflicting behavior attacks. This additional check is based on checking the trust level of the recommenders in providing recommendations which are maintained separately inside a reputation generation system. In [20], R. Chen et al have proposed two mechanisms to consider the trust recommendations, namely the threshold based filtering and the relevance based trust mechanisms. In threshold based filtering, only nodes with recommender trustworthiness higher than a pre-defined threshold are considered and in relevance based trust scheme, only recommenders with high trust in a particular context are taken into considerations. Based on these criteria, selected recommendation trusts are aggregated using the weighted average method before they are combined with the direct trust through another weighting function to form an aggregate trust.

*B. Entropy based Probability Model*

In [10-12, 14], Y. Sun et al. have proposed two models based on the probability theory to govern trust propagation through a third party. The first model is called the trust concatenation model. Suppose node $i$ wants to establish the trust level of node $j$ through a recommender $k$, the trust concatenation model defines the formula in (2) to merge the trust probabilities.

$$P_{i,j} = P_{i,k}P_{k,j} + (1 - P_{i,k})(1 - P_{k,j}) \qquad (2)$$

where $P_{i,j}$ is the probability that node $j$ performs an action, $P_{i,k}$ is the probability that recommender $k$ makes good recommendations and $P_{k,j}$ is the probability that node $j$ performs some actions in the recommender's view. The probability values are estimated by evaluating the expected value of the Beta distribution [25]. The second model is called the multipath propagation model which is used in scenarios where there are multiple recommendation paths between node $i$ and node $j$. In the second model, the mean and variance of the each path $m$ is first converted to Beta parameters $(\alpha_m, \beta_m)$. Then, a new pair of parameters $(\alpha, \beta)$ is updated as follows in (3).

$$\alpha = \sum_{i=0}^{m} \alpha_i \qquad \forall i = 1, \dots, m \qquad (3)$$

$$\beta = \sum_{i=0}^{m} \beta_i \qquad \forall i = 1, \dots, m$$

Once $\alpha$ and $\beta$ values are updated, the aggregated probability is found by evaluating the expected value of the beta distribution. After that, the aggregated probability is converted to trust values using the entropy function as described in (4).

$$T_{i,j} = \begin{cases} 1 - H(P_{i,j}), & for\ 0.5 \leq P_{i,j} \leq 1 \\ H(P_{i,j}) - 1, & for\ 0 \leq P_{i,j} < 0.5 \end{cases} \qquad (4)$$

$$H(P_{i,j}) = -P_{i,j} log_2(P_{i,j}) - (1 - P_{i,j}) log_2(1 - P_{i,j}).$$

## C. Subjective Logic Operators

In [26], Josang have proposed an algebra that is able to quantify uncertainty in trust relationships between entities. Such uncertainties occur due to the imperfect knowledge about the reality or due to the lack of evidence. Consequently, this leads to the notions of belief, disbelief, and uncertainty which forms the basis of subjective logic. In this model, trust is treated as a subjective opinion consisting of four parameters $(b_{i,j}, d_{i,j}, u_{i,j}, a_{i,j})$ where $b_{i,j}$ and $d_{i,j}$ represents node $i$'s belief and disbelief in node $j$, $u_{i,j}$ represents the amount of uncertainties regarding the observations and $a_{i,j}$ is the a prior probability in the absence of evidence. These parameters have the following relationships and satisfy the following conditions in (5).

$$b_{i,j} + d_{i,j} + u_{i,j} = 1.0, \qquad b_{i,j}, d_{i,j}, u_{i,j}, a_{i,j} \in [0.0, 1.0] \qquad (5)$$

From the definition of an opinion, trustworthiness of a node is derived by evaluating the expectation of the subjective opinion as defined in (6)

$$T_{i,j} = b_{i,j} + a_{i,j} u_{i,j} \qquad (6)$$

where $T_{i,j}$ represents node $i$'s trust in $j$ and the parameter $a_{i,j}$ determines how much uncertainty contributes towards the trust computation. Because subjective logic is able to account for uncertainty, it improves the clarity and expressiveness of an opinion compared to the traditional probabilistic logic. The subjective logic further establishes two operations [27, 28] to handle trust propagations in a network. The first operator is called the discounting operator which is used to derive a trust opinion from transitive paths via a recommender. The second operator is called the consensus operator and is used to derive a trust opinion from multiple parallel paths. The concept of subjective logic has been very popular and widely used in Mobile Ad Hoc Networks (MANETs) [8, 17, 18] and WMNs [19] including VANETs [29] with some slight variations. In [17], the trust opinions from multiple recommenders are first combined into a single opinion using the weighted average approach before it is fused with the direct trust using the consensus operator. In [18], the weighting factor for each recommendation trust opinion is derived based on a familiarity value that denotes the familiarity degree of the recommender with the target node to be evaluated. In [19], each of the recommendation trusts are weighted by the trustworthiness of the recommender. Subsequently, the recommendation trusts are aggregated using a simple average function. In [29], subjective logic is proposed as a tool to merge opinions coming from different misbehaviour detection mechanisms together to enhance the trust accuracy and detection capability of the system.

*D. Regression Analysis*

In [15, 22], Y. Wang et al have proposed the LogitTrust model to estimate the trustworthiness of nodes with logit regressions. It gathers the direct and indirect observations as a function of the operational and environmental factors to infer the trustworthiness of a node. Trust, in this model, is expressed as a logistics function and is given by (7).

$$T_j^{t+1} = \frac{1}{1 + e^{-(x^t)^T \beta_j}} \quad (7)$$

where $T_j^{t+1}$ represents the trust of node $j$ at time $t+1$, $x^t$ is a vector of variables that characterize the operational and environmental factors at time $t$ and $\beta_j$ is a vector of regression coefficients. The Expectation Maximization (EM) algorithm is then used to estimate the regression coefficients such that the log-odds of a node in providing a satisfactory service are greater than 0. After that, the regression coefficients are substituted back into (7) to compute the trust of a node. The LogitTrust achieves resiliency against badmouthing and ballot-stuffing attacks by replacing the latent error in logistics distribution with a white noise in t-distribution. Because the t-distribution has heavier tails, the impacts of records with a high variance are weighted down; therefore, it provides a more accurate estimate. However, this method is sensitive to sampling bias and requires more samples to achieve stable and accurate results.

## III. SYSTEM MODEL

### A. Network Model

We consider the Infrastructure-based WMN and the Hybrid-based WMNs in our model. The Infrastructure-based WMN is made up of static mesh routers that establish an infrastructure backbone for the clients. These static mesh routers can be installed within a building to connect stationary computers and other related devices together to form a small-size enterprise network [2]. The mesh routers can also be installed on the rooftop of multiple buildings to bridge communication between them to form an even larger corporate enterprise networking. Some of the mesh routers are equipped with gateway functionality that enables them to provide internet connectivity and to relay traffic to and from the Internet to support the entire enterprise network. This greatly improves the information flow among departments and facilitates management of data. In contrast, the Hybrid based WMNs consist of static mesh routers and mobile clients. The mobile clients roam around the network and can access the network services through the mesh routers or through direct meshing with other mesh clients. An application scenario of Hybrid based WMNs is the Vehicular Networks (VNs) as discussed by [30]. The roadside units are assumed to be the mesh routers and the mobile mesh clients are the vehicles on the road. In this case, WMN enabled VNs can be used to support applications including accident reporting, collision avoidance, electronic toll collection and etc.

*B. Attack Model*

We consider two types of attacks in the system: reputation based attacks that subvert the proper operation of the reputation systems and packet dropping attacks that are due to the selfish behavior of the nodes. Selfish nodes, in this case, try to preserve their own resources while exploiting the services of others and depleting their resources.

First, we consider two reputation based attacks, namely the badmouthing attacks and ballot-stuffing attacks. In the bad mouthing attacks, the recommender node attempts to ruin the reputation of a well-behaved node intentionally by providing bad recommendations against it so as to decrease the chance of that node being selected for service. Ballot-stuffing attack, on the other hand, does the exact opposite. The malicious recommenders attempt to boost the trust of another malicious node by providing good recommendations so as to increase the chance of that malicious node being selected as a forwarder. Subsequently, the trustee may conduct other attacks such as the packet dropping attacks.

Second, we consider two types of packet dropping attacks that are the blackhole and grayhole attacks. By a blackhole attack, a malicious node will advertise itself as having the best route to the destination even though it does not have a route to it. It does this by sending a Route Reply (RREP) packet immediately to the source node [4]. The source node, upon receiving this malicious RREP assumes the route discovery is complete and ignores all other RREPs from the other nodes and selects the path that now includes the malicious node as a relay node to forward the data packets. Subsequently, the malicious node drops all the traffic received to create a blackhole in the network. A grayhole attack is a variation of the blackhole attacks. Instead of dropping all the traffic, the node drops the packets selectively that makes grayhole attacks more difficult to detect as the selective packet dropping may be perceived as packet loss because of the unreliable wireless channels in the networks. Hence, the grayhole attacks may go undetected for a longer period of time than the blackhole attacks [4]. Furthermore, the grayhole nodes may appear well-behaved during the route discovery process or at the starting of data transmission. Subsequently, they will start to drop packets randomly that originate from specific peers in the network.

## IV. PRELIMINARIES

In this section, we introduce the concept of DST [23] and highlight the motivations behind the use of DST that serves as the basis of our proposed DS-Trust model.

### A. Background

DST is a theory of belief based on plausible reasoning and uncertainty [23]. It has two interesting features, one of which is the ability to quantify uncertainty. To illustrate this idea, let us assume that node $X$ has a belief of 0.8 in the trustworthiness of node $Y$ as a forwarding node. If DST is used, the remaining belief of 0.2 is classified as uncertainty because there is no evidence to support that node $Y$ is untrustworthy. With traditional probability theory, the remaining probability is assigned to the untrustworthy state to obey the additive rule of probability. Hence, DST is more flexible than traditional probability theory in handling ignorance or lack of evidence. Another feature of DST is the ability to handle conflicting evidences from multiple sources. This is reflected in the Dempster's rule of combination operator where the conflicting information is ignored through a normalization factor to emphasize the agreement among multiple evidences. These unique features make DST very appealing for modelling trust relationships, including trust aggregation, especially in wireless networks where unreliable overhearing and contradicting opinions because of misbehaviors are prevalent. In the following, we focus on the three important functions related to DST.

### B. Concepts of DST

Let $\Theta$ be the frame of discernment containing a finite set of possible possibilities. Let $2^\Theta$ be a power set of $\Theta$ that contains singleton and all possible unions of the singletons including $\Theta$. An evidence source generates a belief mass or bpa denoted by a mass function $m_j(\cdot)$ for various subsets of the power set where $j = \{1, \dots, J\}$ refers to the index of the evidence source. In this case, the mass function, $m_j(\cdot)$ is a probability defined as a mapping from a set $s\epsilon 2^\Theta$ to a non-negative value between 0 and 1 given in (8).

$$m_j(s): 2^\Theta \rightarrow [0,1], \qquad s \in 2^\Theta \tag{8}$$

$$m_j(\phi) = 0 \tag{9}$$

$$\sum_{s\epsilon 2^\Theta} m_j(s) = 1 \tag{10}$$

The bpa expresses the strength of the evidence pertaining to the subset of $2^\Theta$ under consideration. When assigning bpa, two conditions must be adhered to, that is. no bpa should be assigned to the empty set $\phi$ and the sum of belief mass should be equal to 1 according to (9) and (10) respectively. DST further defines two functions that make use of the bpa. They are called the belief function and plausibility functions. Belief functions are the sum of all bpas that supports a proposition, $s$ defined in (11) while the plausibility function is represented by the sum of all masses that partially or fully support a proposition, $s$ as in (12). Together, the belief and plausibility functions define a belief interval bounded by $[belief(s), plausiblity(s)]$ where the probability of a set $s\epsilon 2^\Theta$ can be obtained.

$$belief_j(s) = \sum_{\substack{p \subseteq s \\ p \in 2^\Theta}} m_j(p) \tag{11}$$

$$plausibility_j(s) = \sum_{\substack{p \cap s \neq \emptyset \\ p \in 2^\Theta}} m(p) \tag{12}$$

C. *Dempster's Rule of Combination*

For a given frame of discernment, it is possible for multiple sources to provide their evidence. Assuming that evidence comes from independent sources, they can be combined in a pairwise manner using the Dempster's rule of combination defined in (13) to arrive at a common shared belief.

$$m_{DS}(s) = m_1(s) \oplus m_2(s) \oplus \ldots \oplus m_J(s) \qquad \forall s \in 2^\Theta \tag{13}$$

where $m_{DS}(s)$ denotes the resulting mass function after combination and $\oplus$ represents the combination operator. If there are only two evidence sources, $E_i$ and $E_j$, the pairwise combination operator is written as follows:

$$m_{i,j}(s) = \frac{1}{1-K} \sum_{\substack{p \cap q = s \\ p \in 2^\Theta \\ q \in 2^\Theta}} m_i(p) m_j(q) \qquad s \neq \phi \tag{14}$$

$$K = \sum_{\substack{p \cap q = \phi \\ p \in 2^\Theta \\ q \in 2^\Theta}} m_i(p) m_j(q) \tag{15}$$

where $m_i(p)$ and $m_j(q)$ represent the bpa assigned to $belief_i$ and $belief_j$ functions respectively. The quantity $K$ denotes the amount of conflict between the two evidence sources and is treated as the normalization factor to ensure that the total sum of combined masses $m_{i,j} = 1$. In essence, the DST's rule of combination sums up all the possible intersections of the propositions and normalized the value by removing all the conflicts in the system.

## V. DS-TRUST MODEL

In this section, we describe the five modules of the DS-Trust model which is shown in Figure 1 and their relationship to each other. The five modules are installed on every node, and they are the monitor module, the feedback module, the correlation module, the fusion module and the decision module.

### A. Monitoring Module

The monitor module is equipped with the Watchdog mechanism [5] to monitor the next hop forwarding behavior by keeping track of the number of packets sent and the number of packets overheard locally. More specifically, each node stores the packet ID and the node ID that the packet is directed to in a table. When each node overhears a packet sent and finds a match in the corresponding table, the table entry corresponding to the overheard packet ID is deleted. During each trust monitoring period denoted as $T$, the node tallies the sum in the table and computes the forwarding probability of the downstream node as in (16). Next, the forwarding probability of a node is converted into a trust value using a set of equations (17) and (18).

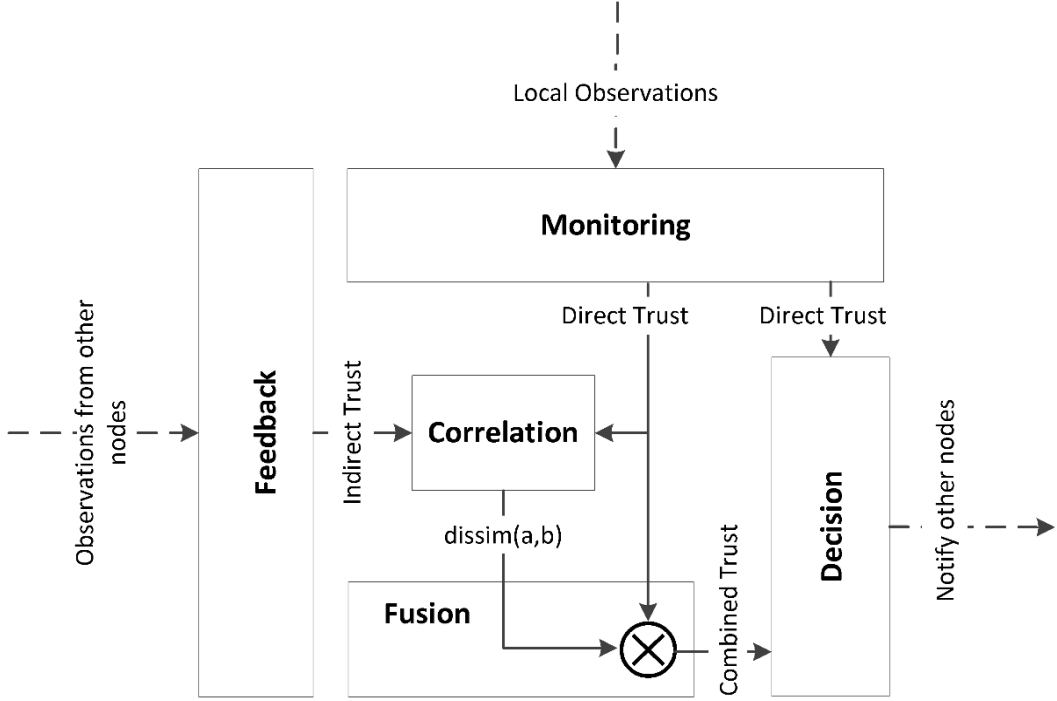

Fig. 1. DS-Trust Model.

$$p_f = \frac{\text{\# of overheard packets sent by } n_i}{\text{\# of packets send to } n_i \text{ for forwarding}} \qquad (16)$$

$$DT = \begin{cases} 1 - 0.5 H_b(p_f), & \text{for } 0.5 \leq p_f \leq 1 \\ 0.5 H_b(p_f), & \text{for } 0 \leq p_f < 0.5 \end{cases} \qquad (17)$$

$$H_b(p_f) = -p_f \log_2 p_f - (1 - p_f) \log_2 (1 - p_f) \qquad (18)$$

where $DT$ denotes the trust value of a node, $H_b(p_f)$ denotes the binary entropy function and $p_f$ denotes the forwarding probability of a node. The use of the entropy function is to reflect the amount of uncertainty about the gathered information because the promiscuous mode of observations is affected by channel conditions and transmission power [5]. The mapping function in (17), on the other hand, is to bind the trust values in the interval (0,1) where a trust value of 1 means a node is fully trusted and a trust value of 0 means a node is not trusted. Furthermore, we incorporate the exponential averaging function as shown in (19) to give more weight to the recent trust value than the past trust values.

$$DT_t = \alpha \cdot DT_t + (1 - \alpha) \cdot DT_{t-1} \tag{19}$$

where $\alpha$ is a constant smoothing factor between 0 and 1, $DT_t$ represents the trust value at time $t$ and $DT_{t-1}$ represents the previous trust value recorded by the monitoring module. If the smoothing factor $\alpha$ is large, it discounts the previous trust faster. Once the direct trust is obtained, it is passed to the correlation module and the decision module that is discussed later.

## B. Feedback Module

This module is responsible for gathering recommendation trusts from the neighboring nodes. During every trust monitoring period $T$, the node broadcasts a request message for the direct trust of a target node. When the neighboring nodes receive the request message, each of them checks its own record to see if there is a trust record for the target node. If a record is found, the neighboring nodes send a recommendation message containing the trust value of the target node back to the requestor. In this case, the recommendation trusts from the recommenders are calculated the same way as described in the monitoring module. When the requestor receives the recommendation message, it weighs the received recommendation trust value by its opinion of the recommender to calculate the indirect trust and then, outputs it to the correlation module. Suppose node $C$ is the target node, node $A$ is the requestor soliciting the opinion of the recommender $B$, the indirect trust denoted by $IDT$ is formulated as (20).

$$IDT_{AC} = DT_{AB} \cdot DT_{BC} \tag{20}$$

## C. Correlation Module

The correlation module and the fusion module are the two core components of the DS-Trust model that help to mitigate badmouthing and ballot-stuffing attacks. Different from the approaches in [7, 9-12, 14], every single recommendation trusts received by the correlation module are considered in the trust aggregation regardless they are good or bad. The correlation module uses the dissimilarity test to compare the direct trust record and the received indirect trust record. The dissimilarity ratio measures the amount of conflict between one's own opinion and the opinions of others. It determines how much the indirect trust records contribute towards the final trust aggregation. The dissimilarity ratio, as shown in (21), is expressed as the normalization of the absolute difference between the two trust records. If the dissimilarity ratio is large, it implies that the two trust records are in conflict. Thus, the evidence supporting the indirect trust record is viewed as uncertainty instead of being filtered out. On the other hand, a small deviation means that the two trust records are almost similar and this amplifies the belief with reduced uncertainty about the observed proposition.

$$dissim(a,b) = \frac{|a-b|}{|a|+|b|} \quad (21)$$

*D. Fusion Module*

The role of the fusion module is to evaluate the trust value of a node by fusing the direct trust and the indirect trust together. Before the actual aggregation takes place, the results of the dissimilarity test are used to re-evaluate the contribution of the indirect trust values and this is carried out based on the Dempster Shafer Theory (DST) [23]. We first classify the behavior of the nodes into two states: trusted ($T$) and untrusted ($\bar{T}$) that forms the frame of discernment $\Theta$ defined as $(T, \bar{T})$. For the power set denoted by $2^\Theta$, it consists of the following subsets:

$$2^\Theta = [(T), (\bar{T}), (T, \bar{T}), (\emptyset)] \quad (22)$$

The set represented by $(T, \bar{T})$ denotes uncertainty in our model that means that a node can be trusted or untrusted. To facilitate the assignment of bpas, we apply the direct trust values obtained from (17) or (19) as the bpas to denote the strength of evidence pertaining to a particular subset of $2^\Theta$. In the design, if the trust value is above a certain detection threshold called $\gamma$, it will be classified as a trusted node, whereas a trust value less than $\gamma$ will be classified as an untrusted node. As an example, if the threshold $\gamma$ is 0.5 and the direct trust value of a node is 0.6, the bpa assigned to the set $m(T)$ will be 0.6. The remaining belief mass of 0.4 will be allocated to the set $m(T,\bar{T})$. For classification of indirect trust values, we leverage on the $dissim(a,b)$ value received from the correlation module and classify the node according to rules defined in (23).

If $IDT \geq \gamma$ then, (23)

$$m(T,\bar{T}) = dissim(a,b)$$

$$m(T) = 1 - m(T,\bar{T})$$

$$m(\bar{T}) = 0$$

Else

$$m(T,\bar{T}) = dissim(a,b)$$

$$m(\bar{T}) = 1 - m(T,\bar{T})$$

$$m(T) = 0$$

From (23), the dissimilarity value is treated as the belief mass for the set $m(T,\bar{T})$ because the amount of conflict between the two trust records denotes uncertainty. Next, depending on the value of the received IDT, the belief mass for the set $m(T)$ and $m(\bar{T})$ is updated accordingly such that $m(T) + m(\bar{T}) + m(T,\bar{T}) = 1$ as stated in (23). Once the belief masses have all been updated for the direct and indirect trust, the Dempster's rule of combination is applied to fuse the two trust evidences together and the result is sent to the decision module for actions. We provide a numerical example to illustrate the trust aggregation process.

Example: Let us suppose that the first evidence $E_1$ which represents node $A$'s direct trust value of node $C$, is 0.9. Because it is more than the detection threshold $\gamma$ of 0.5, a bpa value of 0.9 is assigned to the set $m(T)$ and the remaining 0.1 is assigned to the uncertainty set $m(T,\bar{T})$. Suppose now the IDT about node $C$ from one of the recommenders, say node $B$ is 0.1. The dissimilarity ratio between $A$'s trust and $B$'s trust will be 0.8 according to (21). Thus, for the second set of evidence, $E_2$ that represents the indirect trust value of node $C$ based on $B$'s recommendation, $m(\bar{T})$ will be assigned with 0.2 since IDT is less than $\gamma$ and the remaining 0.8 will be allocated to set $m(T, \bar{T})$. With this information, we tabulate Table I and use the Dempster's rule of combination in (14) and (15) to evaluate the final trust. Using the Dempster's rule of combination, the combined belief $m_{1,2}(T)$ that node $C$ is trusted from node $A$'s perspective is

$$m_{1,2}(T) = \frac{1}{1-K}[m_1(T) \cdot m_2(T, \bar{T})]$$

$$= \frac{1}{1-0.18}[0.72]$$

$$= 0.87805$$

| $E_1$ \ $E_2$ | $\{\phi\} = 0$ | $\{T\} = 0$ | $\{\bar{T}\} = 0.2$ | $\{T, \bar{T}\} = 0.8$ |
|---|---|---|---|---|
| $\{\phi\} = 0$ | $\{\phi\} = 0$ | $\{T\} = 0$ | $\{\bar{T}\} = 0$ | $\{T, \bar{T}\} = 0$ |
| $\{T\} = 0.9$ | $\{T\} = 0$ | $\{T\} = 0$ | $\{\phi\} = 0.18$ | $\{T\} = 0.72$ |
| $\{\bar{T}\} = 0$ | $\{\bar{T}\} = 0$ | $\{\phi\} = 0$ | $\{\bar{T}\} = 0$ | $\{\bar{T}\} = 0$ |
| $\{T, \bar{T}\} = 0.1$ | $\{T, \bar{T}\} = 0$ | $\{T\} = 0$ | $\{\bar{T}\} = 0.02$ | $\{T, \bar{T}\} = 0.08$ |

Table I: Aggregation of $E_1$ and $E_2$

*E. Decision Module*

If the network is sparse and there are no recommenders to provide indirect trust, the decision module will base its decision only on the direct trust from the monitoring module. Otherwise, the combined trust value as calculated per the fusion module is used. Below summarizes the actions undertaken by the evaluating node when the trust value falls below the detection threshold $\gamma$.

- Isolate selfish nodes – the evaluating node will blacklist the misbehaved node. At the same time, it conducts a blacklist broadcast throughout the network to inform other nodes who will further block it from all subsequent communications. We wish to highlight that excluding the blacklist nodes permanently from routing is better than assigning low ratings to them and allowing them to regain their trust slowly. It is because the latter may induce the misbehaved nodes to misbehave on and off intermittently to avoid detection which is more challenging to solve. In our model, the blacklist nodes who wish to re-join the network can contact an authorized WMN operator to be reinstated. The WMN operator will track the number of times a particular node has been blacklisted. If a particular node has been blacklisted repeatedly beyond a certain number of counts pre-defined by the WMN operator, it can never re-join the network. By relying on a trusted authority, the WMN operator has a better visibility of the health of the nodes in the network.

- Initiate new route discovery – decision about the trust level of a node is also sent to the underlying routing protocol. It will trigger the routing protocol to send a route error (RERR) message to notify the source node to initiate a new route discovery to find a path free of selfish nodes.

## VI. SECURITY ANALYSIS

We conduct an experiment to analyze numerically the changes in the trust value as a function of increasing badmouthing and ballot-stuffing attackers and compare our results to the benchmarking schemes discussed in the related works. The benchmarking schemes used are the linear opinion pooling technique, entropy-based probability model, subjective logic and the regression analysis technique based on the work of [20], [11], [29] and [22] respectively. The aim of this comparative study is to validate the effectiveness of the existing trust aggregation schemes in comparison to using DST for mitigation of badmouthing and ballot-stuffing attacks. In our comparison with the benchmarking schemes, the main focus is the idea behind each of the various aggregation techniques put forth in the related papers. As such, in the implementation of the linear opinion pooling technique presented in [20], we are not concerned about the optimal weight selection and therefore, have configured the weights for both the direct and indirect trust components of the linear combination function to 0.5. Also in [22], the key point is to assess the effectiveness of using the subjective logic operators to combine trust, instead of validating the fusion of node-centric and data-centric opinions from different detection mechanism as presented in the paper. Nevertheless, we try to model as closely as possible in accordance with each paper to ensure a fair comparison.

*A. Experiment Setup*

We consider the scenario as shown in Figure 2 to illustrate the aggregation of direct trust and indirect trust. We are interested in the number of badmouthing and ballot-stuffing attackers that can swing the trust values into the untrusted region. For this, we assume that the trust detection threshold is set to 0.5. According to Figure 2, node $A$ has local observations about node $B$ that is known as the direct trust. Besides that, there are up to 20 recommenders (node $C\ to\ V$) from which node $A$ can gather recommendation trusts about node $B$. The direct trusts are denoted by solid lines, whereas the recommendation trusts from recommenders $C, D$ etc. are denoted by dotted lines. When node $A$ receives the recommendation trusts, it calculates the indirect trusts by weighing the recommendation trust values based on the trust level of the recommenders $C, D, …, V$ etc. We assume that node $A$ has the same direct trust value on each of the recommenders so that any observed changes in the trust aggregation results are due to the recommendation trust values. In the first experiment, we start by configuring one recommender to send a low rating of 0.1 to node $A$ to simulate badmouthing attacks and increase the number of badmouthing recommenders each round until it reaches 20. The same rule applies to the second experiment except that each recommender is now configured to feedback a rating of 0.9 to simulate ballot-stuffing attacks. In this comparison, the trust value is defined as a continuous value in the range $(0,1)$. Therefore, the trust value of the entropy-based probability model is remapped into the range $(0,1)$ using equation (17) for fair comparisons.

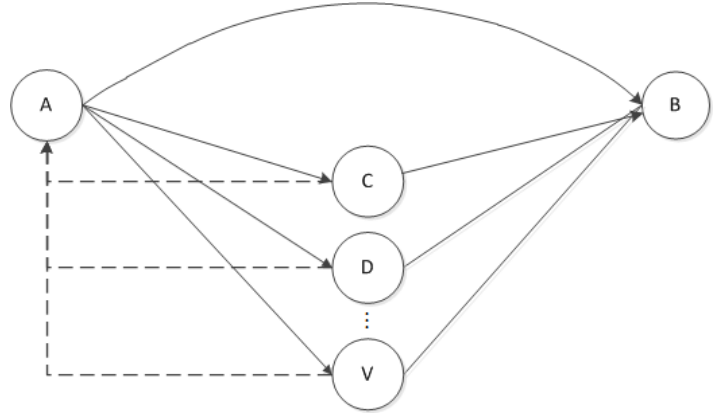

Fig. 2. Trust aggregation.

*B. Results and Discussions*

Figure 3 describes the changes in the aggregate trust value by varying the number of badmouthing attackers. As seen in Figure 3, the trust value of the LogitTrust model, the subjective logic model and the linear opinion pooling technique drops drastically down into the untrusted region when there is only one badmouthing attacker in the network. The entropy-based probability model is slightly better as it is able to tolerate up to two badmouthing attackers. On the other hand, the aggregate trust value of the DS-Trust model has the best performance. It is able to maintain a high trust within the trusted region for up to ten badmouthing attackers before the trust value enters the untrusted region. The improvement is due to the treatment of uncertainty when conflicting trust records are received. In particular, the dissimilarity ratio of the two records is treated as uncertainty that is viewed as either trusted or untrusted in DST framework. Subsequently, when the Dempster's rule of combination is applied to combine the direct trust and the indirect trust, uncertainty is absorbed into the aggregation process that amplifies the belief that the node is trusted. Therefore, the DS-Trust model has a slower trust decay compared to the other benchmarking schemes. This shows that the proposed dissimilarity test and the Dempster's rule of combination are able to mitigate the effects of badmouthing attacks effectively. From Figure 3, we further observe that the aggregated trust value based on the linear opinion pooling technique does not span the full range of possible trust values as seen by the low dip just below the trust threshold boundary of 0.5. The aggregated trust value is 0.495 indicating that the node is untrustworthy. This is due to the choice of the weighting factors in the linear function which we configure as 0.5 in this comparative study. We note that the performance of the linear opinion pooling technique could be improved by using a heavier weight on the direct trust component of the linear function. However, relying too much on the direct trusts would downplay the influence of the indirect trusts towards the final trust aggregation. To resolve this problem, R. Chen et al. [20] has proposed to set the weights dynamically in response to past node changes and environmental changes. However, this approach is not indicative enough to model the actual behavior of a node and can only be regarded as an estimate. The other reason for the small trust variation as seen from the flat curve is that all the indirect trust values are first combined into a single aggregate indirect trust value using the weighted average approach before it is merged with the direct trust value. Therefore, the impact of the aggregated indirect trust is very small in comparison to direct trust observations. Figure 4 shows the trust relationship when there are ballot-stuffing attackers in the network. Similar results are observed where the DS-Trust model is able to tolerate up to ten ballot-stuffing attackers before it succumbs to the false recommendations while the rest of the scheme are vulnerable to a small number of ballot-stuffing attackers. In figure 4, the aggregated trust value based on the linear opinion technique is 0.5 that implies that node *B* is trustworthy. This value is calculated based on the assumptions that the trustworthiness of each recommender from node *A*'s perspective is 1. We reiterate that the linear opinion approach is not effective even through the weights of the linear combination function can be dynamically adjusted according to the operation profile of a node [20]. This is because a node with healthy energy level and higher cooperative index does not necessarily mean that it is cooperative in nature and would adhere to the rules of the protocol.

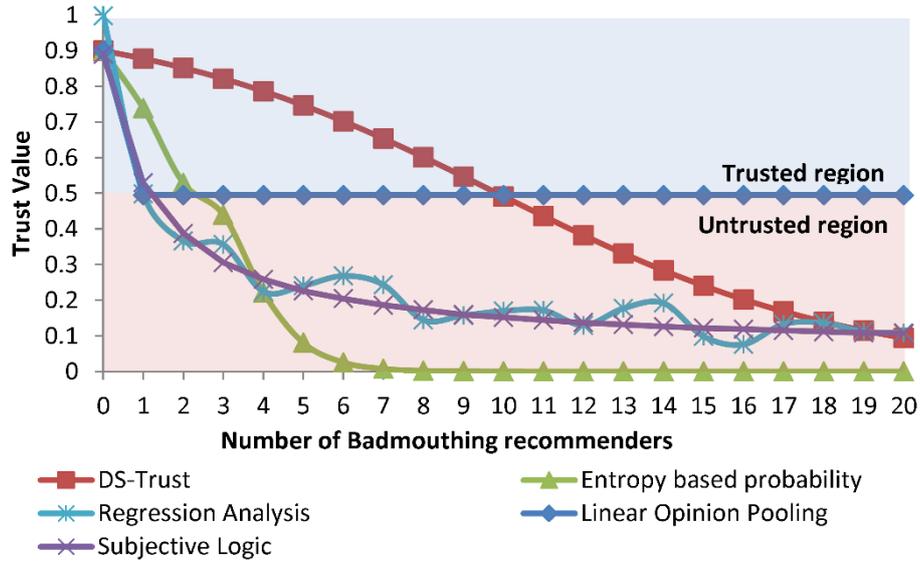

Fig. 3. Trust value as a function of badmouthing recommenders.

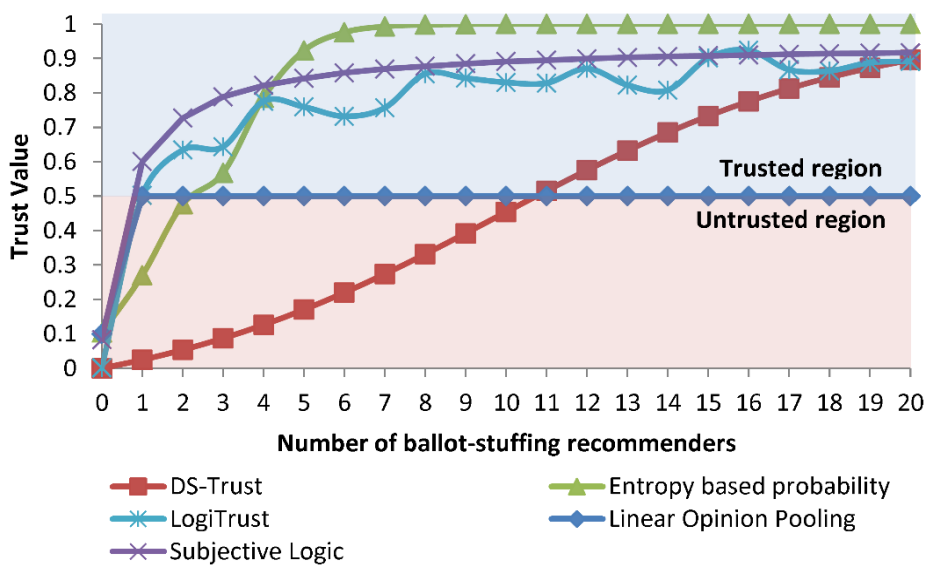

Fig. 4. Trust value as a function of ballot-stuffing recommenders.

## VII. PERFORMANCE ANALYSIS

In this section, we incorporate the DS-Trust model into the AODV protocol [24] and evaluate the performance in terms of the Packet Delivery Ratio (PDR), Normalized Routing Overhead (NRO) and the network throughput. The performance is evaluated under the blackhole and grayhole attacks for two different types of WMN architecture: an Infrastructure based WMN and Hybrid based WMN. The simulations are performed using the Network Simulator NS3 (v3.20) [31] and the results are compared to the baseline AODV and a variant of the DS-Trust that does not consider recommendation trusts at all which we call the DS-Trust (w/o recommendations) model. Lastly, we analyze the computational complexity to evaluate the execution time of our model.

*A. Simulation Environment*

We consider two simulation topologies for the performance evaluation of DS-Trust. The first topology is a static environment to mimic the Infrastructure based WMN while the second topology is a hybrid environment consisting of static and mobile nodes to model the Hybrid based WMN. For the static topology, the source and destination nodes are located on the leftmost and rightmost side of the square grid denoted by a darker color as shown in Figure 5. In the Hybrid based WMN environment as shown in Figure 6, 50 mobile nodes are added to the static topology of 100 nodes and we simulate 8 CBR flows from the four gateway nodes located at four corners of the static grid to any random mobile nodes and vice-versa. The starting time of each flow is uniformly distributed between 30 seconds and 200 seconds. The following performance metrics are used to evaluate the proposed DS-Trust model and the rest of the simulation parameters are given in Table II.

- Packet Delivery Ratio (PDR) refers to the ratio of the number of delivered packets to the number of packets generated by the CBR sources.

- Normalized Routing Overhead (NRO) refers to the number of routing control packets, such as the RREQ, RREP, RERR and the trust related control packets transmitted per data packet delivered at the destination.

- Throughput refers to the amount of data successfully delivered to the intended destinations over a wireless channel. It is measured in bits per second (bps).

- False positive rate is the ratio of the number of nodes that the DS-Trust model misreports as misbehaving to the total number of nodes in the network.

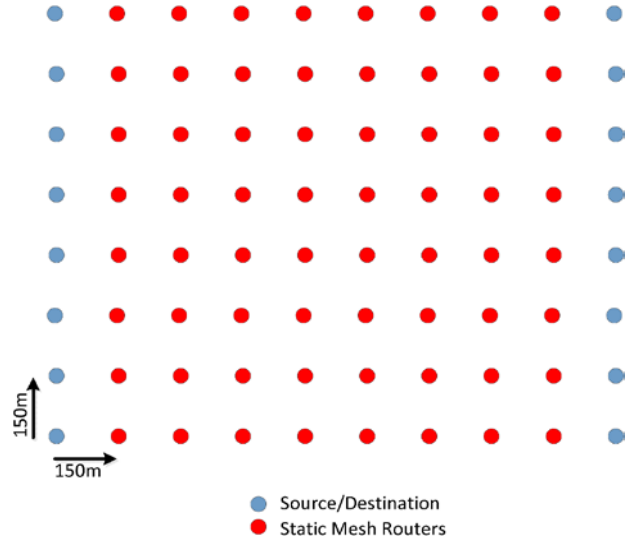

Fig. 5. Simulation topology for Infrastructure based WMN.

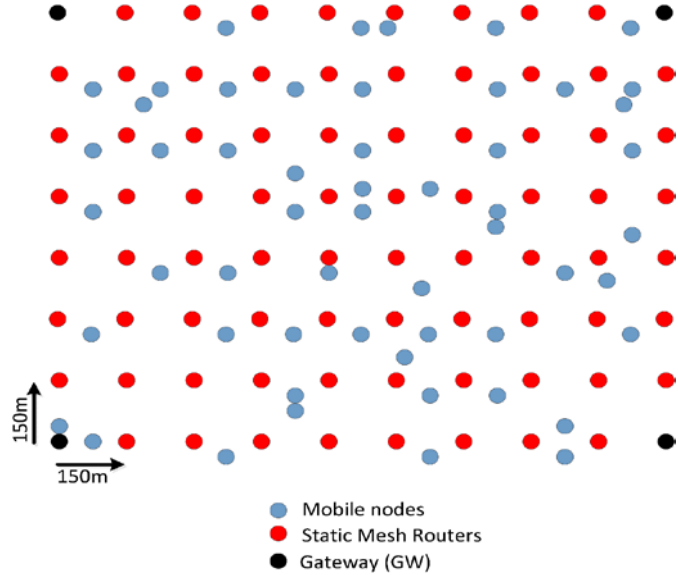

Fig. 6. Simulation topology for Hybrid based WMN.

| Simulation tool | NS-3 |
|---|---|
| **Grid spacing** | 150m |
| **Transmission range** | 250 m |
| **Network area** | 1350 m x 1350 m |
| **Data rate** | 16kbps |
| **Packet size** | 512 Bytes |
| **Packet generation rate** | 4 packets/s |
| **Simulation time** | 300 s |
| **Traffic type** | CBR |
| **Transport protocol** | UDP |
| **Mac protocol** | IEEE 802.11b |
| **Propagation loss model** | RangePropagationLossModel |
| **Physical layer** | YansWifiPhy channel |
| **Detection Threshold, $\gamma$** | 0.5 |
| **Trust Monitoring Period, $T$** | 20s |

| Infrastructure based WMN | |
|---|---|
| Mobility | Static |
| No: of nodes \ | 100 |
| Traffic | 10 source-destination pairs |
| Routing protocol | AODV (disable HELLO) |
| **Hybrid based WMN** | |
| Mobility/Mobility pattern | 50 mobile nodes RandomWaypoint |
| No: of nodes | 100 static 50 mobile nodes |
| Traffic | 8 source-destination pairs |
| Routing protocol | AODV with HELLO |

Table II: Simulation Parameters

## B. PDR Performance

Figure 7 and 8 show the PDR performance of the various schemes in an Infrastructure based WMN and a Hybrid based WMN respectively when the network is under blackhole attacks. From Figure 7, we observe that the PDR of all the schemes remains almost the same when there is no blackhole attacker in the network. As the number of blackhole nodes increases, the PDR starts to decline. However, the DS-Trust model is able to achieve about 30% improvement in the PDR over that of the baseline AODV and about 15% improvement in the PDR over the DS-Trust (w/o recommendations) model. DS-Trust is more superior to the DS-Trust (w/o recommendations) because it gathers recommendation trusts from the neighboring nodes, which improves the detection time of selfish nodes in the network. As the number of blackhole nodes increases further, the PDR of the DS-Trust model starts to decline gradually because of more blackhole nodes being identified and isolated, and fewer alternatives are available for the forwarding paths. Similar results can be observed in the Hybrid based WMN shown in Figure 8 where the PDR improvement for the DS-Trust model is about 14% higher and 10% better compared to the baseline AODV and DS-Trust (w/o recommendations) model respectively. The PDR improvement is lower for the Hybrid based WMN than in the Infrastructure based WMN because it is more difficult to maintain link stability because of the node movement. Moreover, the effect of not using the recommendation trusts from the other nodes is more visible in this plot as there is almost no improvement in the PDR when compared to the baseline AODV.

Next, we study the PDR performance of the Infrastructure based WMN and the Hybrid based WMN under the influence of grayhole attackers. We assume that 10% of the network nodes are grayhole attackers and they perform dropping rates between 0% and 100%. Simulation results in Figure 9 show that the PDR performance of the DS-Trust model is almost similar to the baseline AODV when the selective dropping probability is between 0 and 0.4. This is expected because the trust detection threshold for the DS-Trust model is configured as 0.5. When the grayhole nodes start to drop the packets at a rate of 50% or more, the DS-Trust model outperforms the baseline AODV and the DS-Trust (w/o recommendations) model. Furthermore, the DS-Trust model improves the PDR to an average of about of 80%. On the other hand, the DS-Trust (w/o recommendations) model could only improve the PDR by an average of 15% because it does not rely on the recommendation trusts for trust evaluations. Similar results are observed in the Hybrid based WMN topology. As shown in Figure 10, the PDR of the DS-Trust model improves only when the grayhole attackers exhibit 50% or more dropping rate, which corresponds to the trust detection threshold in our model.

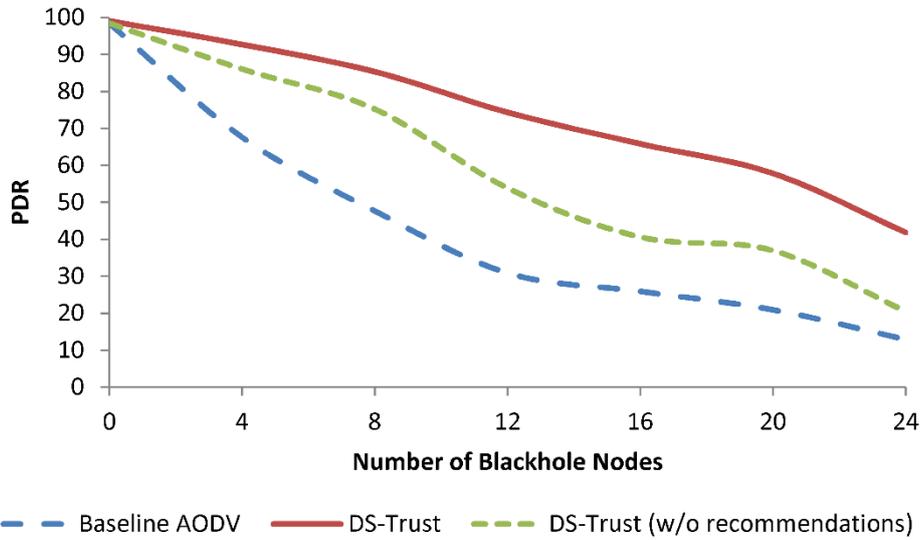

Fig. 7. PDR performance in an Infrastructure based WMN

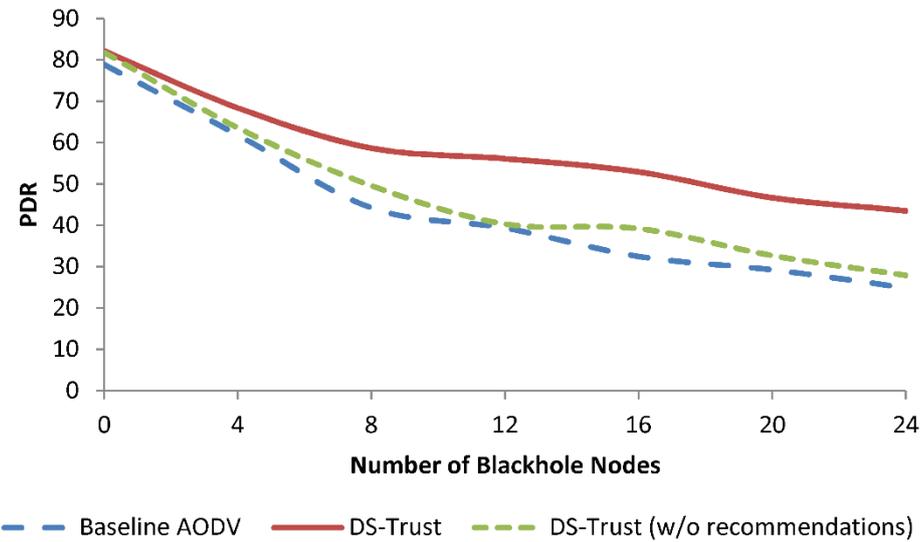

Fig. 8. PDR performance in a Hybrid based WMN

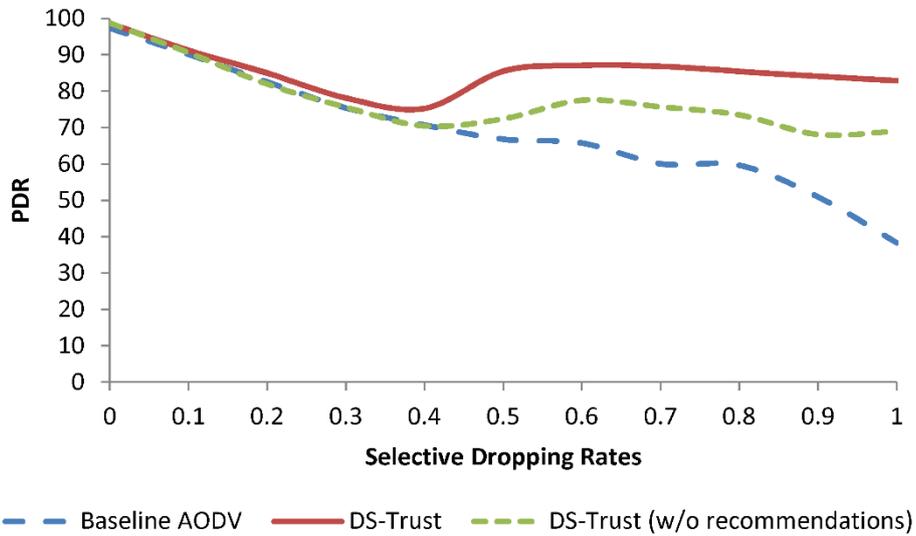

Fig. 9. PDR performance in an Infrastructure based WMN in the presence of 10% grayhole

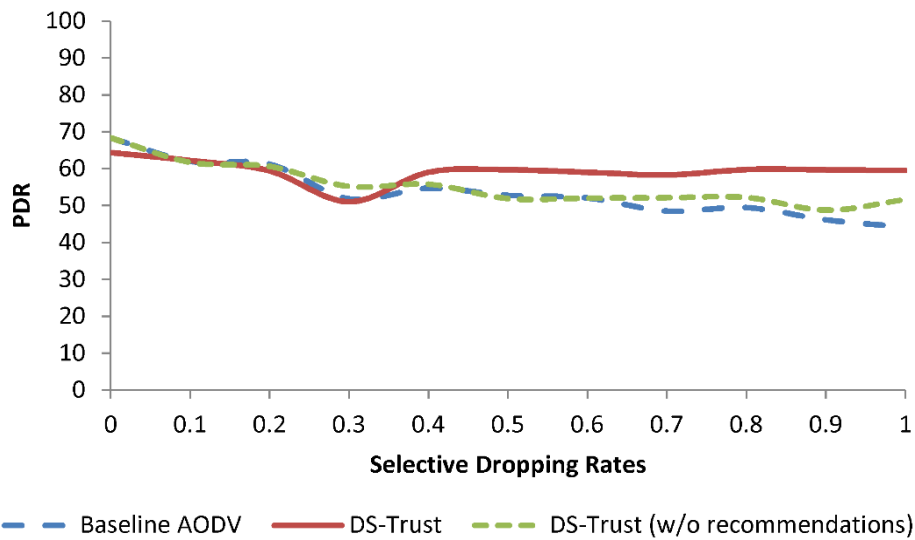

Fig. 10. PDR performance in a Hybrid based WMN in the presence of 10% grayhole

However, the PDR improvement is only 10% better than the baseline AODV compared to the 28% in the static case. Similarly, the PDR improvement over the DS-Trust (w/o recommendations) model is only about 7%-8% compared to 12%-13% in the static environment. The main reason for the lower PDR improvement is node mobility. When nodes are mobile, the links become relatively unstable and less reliable that results in more packets drop. In the case of DS-Trust (w/o recommendations) model, the PDR did not improve at all because the recommendation trusts are not disseminated to other nodes to allow them to make a better judgment.

## C. NRO Performance

The NRO metric is a measure of the effective use of the wireless channel. First, we examine the NRO performance under the blackhole attacks. Next, we analyze the performance in the presence of grayhole attacks. Figure 11 compares the NRO performance of the various schemes in an Infrastructure based WMN. Simulation results show that the baseline AODV has the lowest NRO followed by the DS-Trust (w/o recommendations) model and the DS-Trust model. The NRO performance of the DS-Trust model is higher than the DS-Trust (w/o recommendations) model because of the dissemination of recommendation trusts. In addition, both the DS-Trust models are higher than the baseline AODV because of the extra control packets introduced by our trust model, especially the periodic exchanges of trust information, the broadcast of control messages and the re-initiation of new route discoveries upon detection of blackhole nodes. On the other hand, in the Hybrid based WMN scenario, as shown in Figure 12, the NRO performance of the DS-Trust model is much lower than the baseline AODV compared to the static case in Figure 11. This is because when a secure path is found, it is unlikely to change unless the node moves out of transmission range. Even when the link breaks because of mobility, the list of blackhole nodes is circulated to the other neighboring nodes. Therefore, the routing protocol avoids them during a new route discovery that leads to lesser route discoveries and lesser NRO.

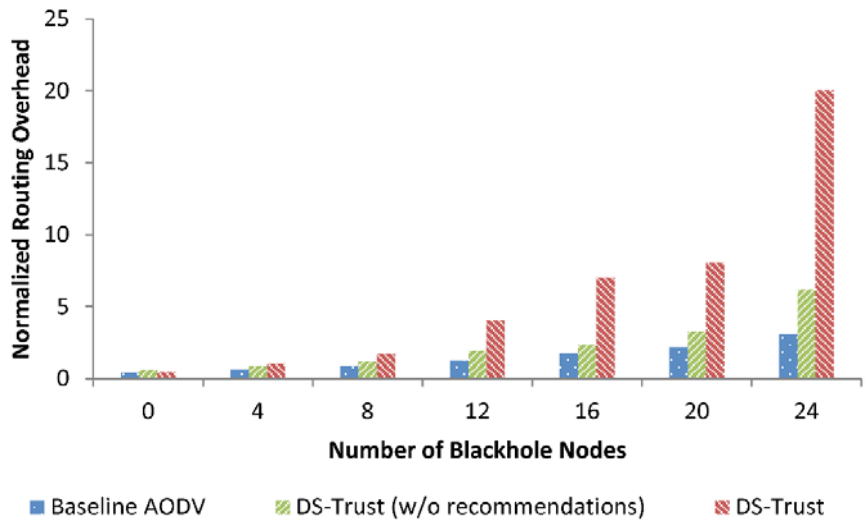

Fig. 11. Normalized routing overhead performance in an Infrastructure based WMN

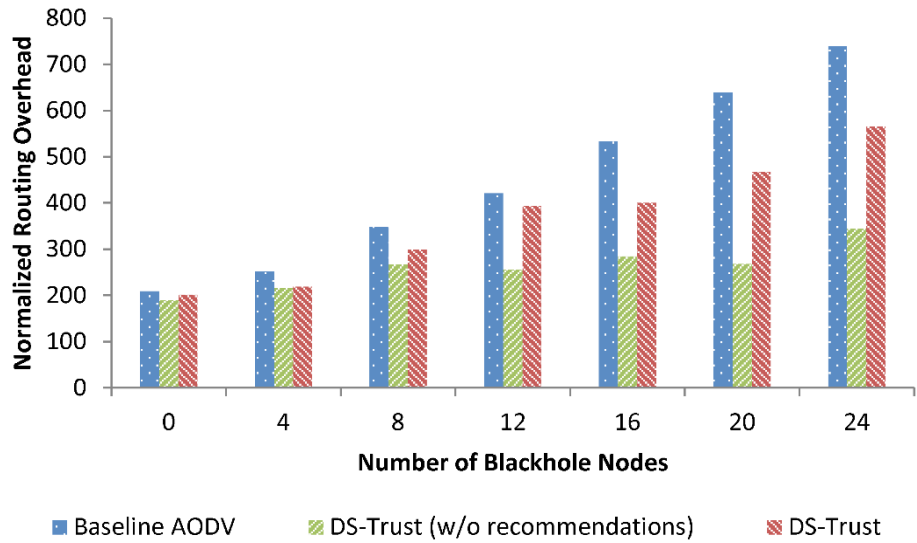

Fig. 12. Normalised routing overhead performance in a Hybrid based WMN

In the case of selective dropping or grayhole nodes in an Infrastructure based WMN, the increase in the NRO is more apparent when the dropping rate is 50% and more as depicted in Figure 13. This is true because the trust detection threshold is 0.5. As a result, more control packets are being broadcast to inform the other nodes of the grayhole nodes, including the control packets needed for a new route discovery. On the other hand, the NRO performance for both the DS-Trust models in the Hybrid based WMN are quite similar to the Figure 12 where both the NRO performances are lower than the baseline AODV. This is illustrated in Figure 14.

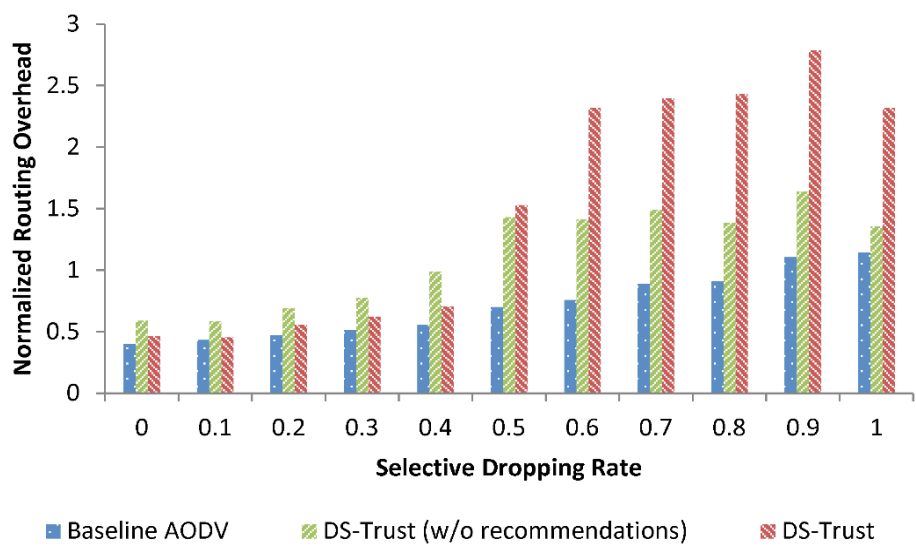

Fig. 13. Normalized routing overhead performance in an Infrastructure based WMN in the presence of 10% grayhole

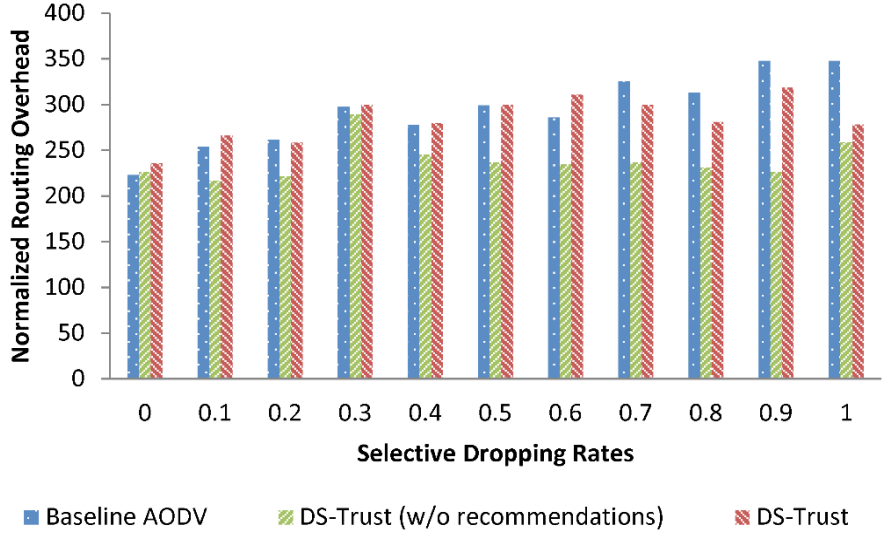

Fig. 14. Normalized routing overhead performance in a Hybrid based WMN in the presence of 10% grayhole

## D. Effects of Mobility

Next, we investigate the effects of mobility on the PDR and the NRO performance in a Hybrid based WMN under the influence of the blackhole nodes. The number of blackhole nodes in the network is 8 and the speed is varied from 10m/s to 40m/s. Simulation results in Figure 15 show that the DS-Trust model has the highest PDR performance compared to the baseline AODV and the DS-Trust (w/o recommendations) model. As the node speed increases, the PDR decreases for all the schemes because of more link breakages in the network. Figure 16 illustrates the NRO performance for different mobility speeds. It is observed that the NRO performance increases when the node mobility is high. However, the NRO performance of the DS-Trust model is lower than that of the baseline AODV. This confirms our observations that DS-Trust model is able to isolate selfish nodes and exclude them from routing.

## E. Throughput Performance

We are interested in the throughput performance of the DS-Trust model when the application rate is increased in an Infrastructure based WMN and a Hybrid based WMN. We assume there are 8 blackhole nodes and the application rate is varied from 16384bps to 200000bps. Figure 17 compares the throughput performance of the DS-Trust model in an Infrastructure based WMN with the baseline AODV and DS-Trust (w/o recommendations) model. As shown in Figure 17, the DS-Trust model denoted by the line with the 95% confidence tick marks, is able to achieve a higher throughput than the DS-Trust (w/o recommendations) for application rate between 16384bps and 65536bps. Beyond the rate of 65536bps, there is no performance improvement between the two models. This could be due to the blacklist of false positives located near the source node that results in no available forwarding paths to send the packets to the destination. We also observe that the throughput for the DS-Trust and the DS-Trust (w/o recommendations) starts to decrease beyond 65536bps. This is caused by the high packet collisions due to the increased sending rate. Because a fixed trust detection threshold is assumed, the DS-Trust models are not able to differentiate losses due to the collisions or malicious intent. Thus, higher false positives are generated, resulting in a reduction in the throughput performance. The throughput performance of the Hybrid based WMN is presented in Figure 18 where it is observed that the DS-Trust model is able to maintain a high throughput throughout the range of the packet sending rates. Comparing the throughput performance of the baseline AODV and the DS-Trust (w/o recommendations) model, the throughput improvement of the DS-Trust is about 25% higher. The throughput of the DS-Trust (w/o recommendations) model remains low because the detection of selfish nodes takes a long time since the recommendation trusts are not used, hence, more packet drop. Furthermore, blackhole nodes may move to other parts of the network to conduct packet dropping attacks. For the false positive rate, DS-Trust model performs better than the DS-Trust (w/o trust recommendations) model because the trust evaluation is improved using recommendation opinions from many other nodes.

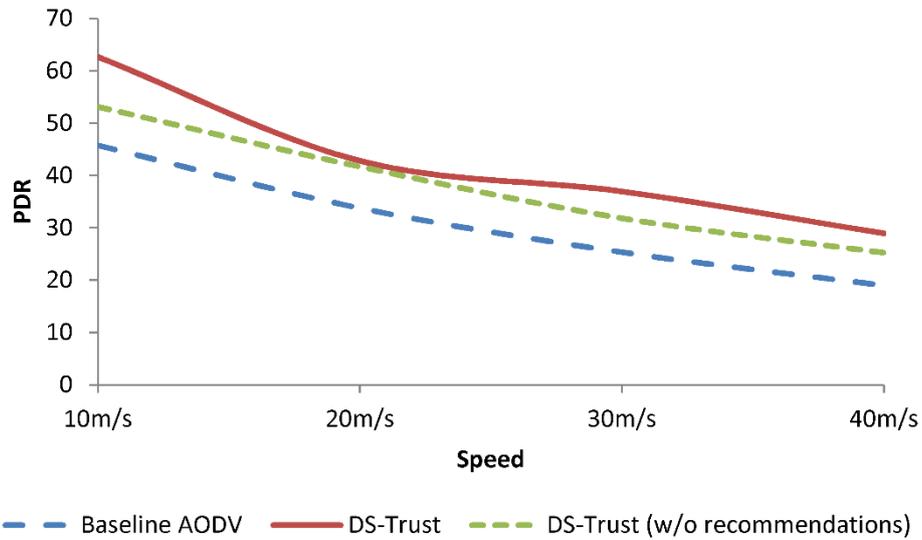

Fig. 15. PDR performance in a Hybrid based WMN under varying speed

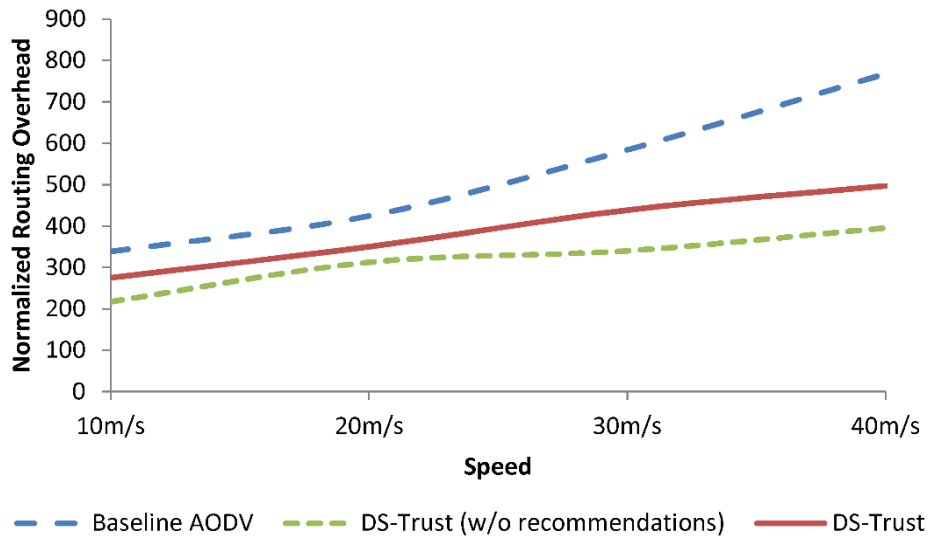

Fig. 16. Normalized routing overhead performance in a Hybrid based WMN under varying speed

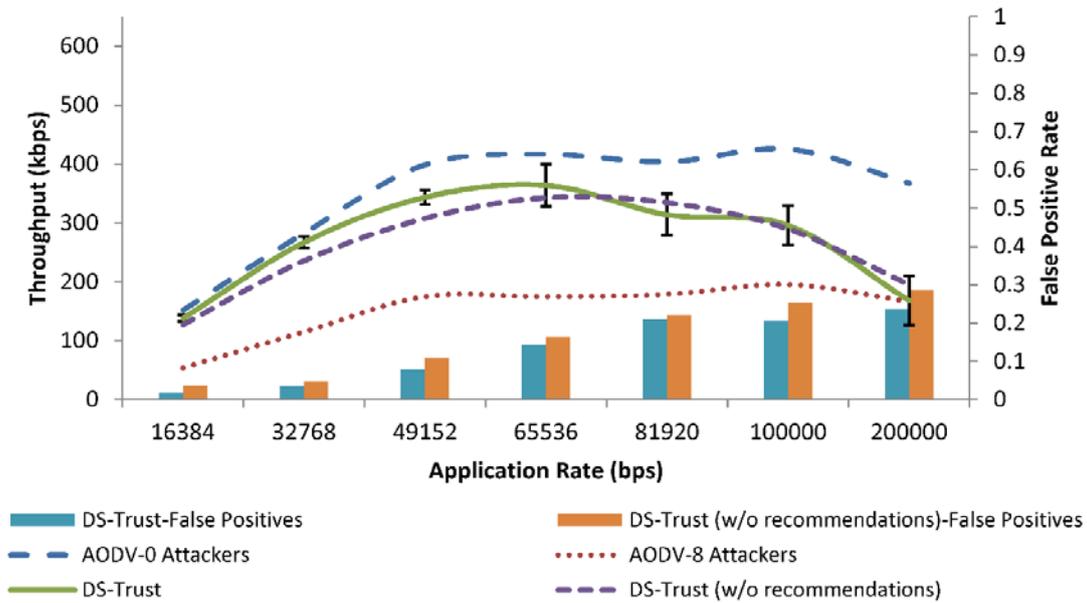

Fig. 17. Throughput Performance in an Infrastructure based WMN

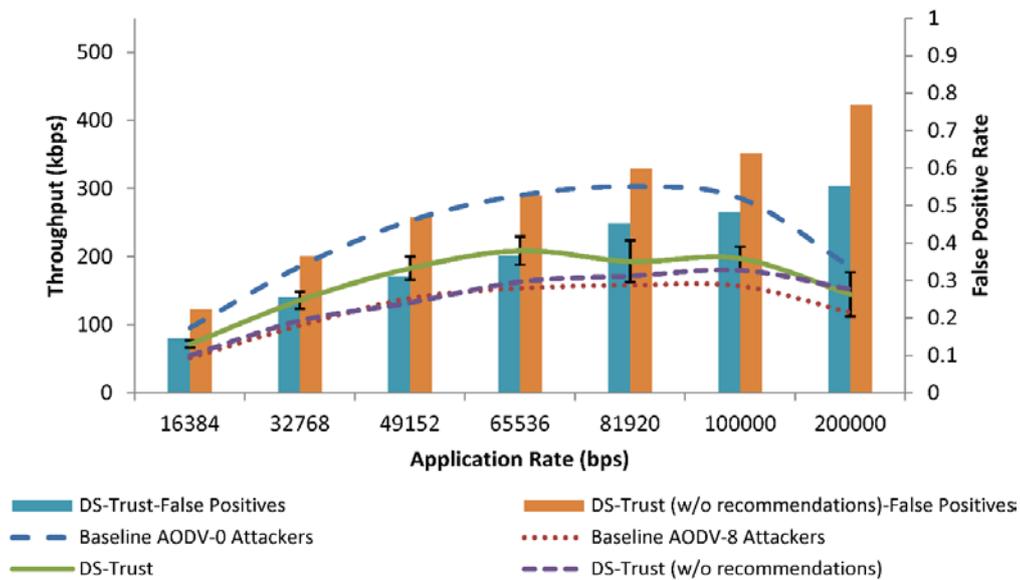

Fig. 18. Throughput performance in a Hybrid based WMN

*F. Computational Complexity*

First, we estimate the computational complexity of each module in the DS-Trust model using the Big O notation. After that, we merge them to determine the overall execution time of the DS-Trust model. In the monitoring module, each node needs to compute the direct trust of the downstream nodes it encounters every trust monitoring period. So the computation complexity of the monitoring module is $O(N)$ where $N$ denotes the number of nodes in the network. The feedback module sends out recommendation requests to solicit recommendation trusts from its one hop neighbors. Upon receiving the recommendation trusts, each node needs to lookup the trust value of the recommender which is $O(1)$ in complexity and weigh each of the received recommendation trusts to calculate the indirect trusts. Suppose there are $n$ recommenders, the complexity of the feedback model is thus $O(n)$. The correlation module takes the direct trust and indirect trust values as input to compute the dissimilarity test. Therefore, the correlation module requires $O(Nn)$ in complexity where $N$ denotes the number of nodes in the network and $n$ is the number of recommenders providing feedback. Next, the fusion module uses the Dempster's rule of combination to fuse two trust records together that is, the direct trust and the indirect trust. The complexity is given by $O(2^S)$ where $S$ is the number of elements in the frame of discernment and $2^S$ corresponds to the number of interactions of the two mass functions. When there are $N$ nodes and $n$ recommenders, the overall complexity of the fusion module is given by $O(Nn * 2^S)$. With these values, the total runtime of the entire trust model is estimated as $O(N) + O(n) + O(Nn) + O(Nn * 2^S)$ where the complexity is dominated by the fusion module. However, we only consider two elements in the frame of discernment. Hence, the complexity of the DS-Trust is low which is $O(4Nn) \approx O(Nn)$. Furthermore, the trust computations are carried out every trust monitoring period $T$ which simplifies the complexity of the whole model as $O(Nn/T)$.

## VIII. CONCLUSIONS

In this paper, we have proposed the DS-Trust model which consists of five modules: a monitoring module, a feedback module, a correlation module, fusion module and a decision module. The monitoring module monitors the next hop forwarding promiscuously and formulates the direct trust using the entropy function to describe the unreliability of promiscuous listening. The correlation module performs dissimilarity test between the direct trust and all the received recommendation trusts to determine the amount of conflict in the trust records. The fusion module then uses the results of the correlation module to re-evaluate the contribution of the indirect trust value and proposes the Dempster's rule of combination to fuse the direct trust and indirect trust together. We have demonstrated numerically that the DS-Trust model is capable of handling highly misleading trust information and mitigate the effects of badmouthing and ballot-stuffing attacks compared to the linear opinion pooling, subjective logic model, entropy-based probability model and logit regression approaches. In addition, we have applied the DS-Trust model to two different WMN architectures and perform extensive NS-3 simulations to demonstrate that DS-Trust is resilient to packet dropping attacks and is able to recover from the blackhole and grayhole attacks. More specifically, DS-Trust is able to improve the PDR and the throughput of the network with reasonably routing overhead. As a future work, we plan to estimate wireless losses due to bad wireless channel quality or medium access collisions to set the trust detection threshold adaptively. This would allow us to detect grayhole attackers with much higher accuracy and further improve the packet delivery ratio and throughput.

## References

[1]  I. F. Akyildiz and W. Xudong, "A survey on wireless mesh networks," *Communications Magazine, IEEE,* vol. 43, pp. S23-S30, 2005.
[2]  I. F. Akyildiz, X. Wang, and W. Wang, "Wireless mesh networks: a survey," *Computer Networks,* vol. 47, pp. 445-487, 3/15/ 2005.
[3]  S. Glass, M. Portmann, and V. Muthukkumarasamy, "Securing Wireless Mesh Networks," *Internet Computing, IEEE,* vol. 12, pp. 30-36, 2008.
[4]  J. Sen, "Secure routing in wireless mesh networks," *arXiv preprint arXiv:1102.1226,* 2011.
[5]  S. Marti, T. J. Giuli, K. Lai, and M. Baker, "Mitigating routing misbehavior in mobile ad hoc networks," in *Proceedings of the 6th annual international conference on Mobile computing and networking*, 2000, pp. 255-265.
[6]  S. Buchegger and J.-Y. Le Boudec, "Nodes bearing grudges: Towards routing security, fairness, and robustness in mobile ad hoc networks," in *Parallel, Distributed and Network-based Processing, 2002. Proceedings. 10th Euromicro Workshop on*, 2002, pp. 403-410.
[7]  P. Michiardi and R. Molva, "Core: a collaborative reputation mechanism to enforce node cooperation in mobile ad hoc networks," in *Advanced Communications and Multimedia Security*, ed: Springer, 2002, pp. 107-121.
[8]  X. Li, M. R. Lyu, and J. Liu, "A trust model based routing protocol for secure ad hoc networks," in *Aerospace Conference, 2004. Proceedings. 2004 IEEE*, 2004, pp. 1286-1295.
[9]  S. Buchegger and J.-Y. Le Boudec, "A robust reputation system for peer-to-peer and mobile ad-hoc networks," in *P2PEcon 2004*, 2004.
[10] Y. L. Sun, Z. Han, W. Yu, and K. Liu, "Attacks on trust evaluation in distributed networks," in *Information Sciences and Systems, 2006 40th Annual Conference on*, 2006, pp. 1461-1466.
[11] Y. L. Sun, Y. Wei, H. Zhu, and K. J. R. Liu, "Information theoretic framework of trust modeling and evaluation for ad hoc networks," *Selected Areas in Communications, IEEE Journal on,* vol. 24, pp. 305-317, 2006.
[12] Y. L. Sun, H. Zhu, Y. Wei, and K. J. R. Liu, "A trust evaluation framework in distributed networks: Vulnerability analysis and defense against attacks," in *INFOCOM 2006. 25th IEEE International Conference on Computer Communications. Proceedings*, 2006, pp. 1-13.
[13] S. Ganeriwal, L. K. Balzano, and M. B. Srivastava, "Reputation-based framework for high integrity sensor networks," *ACM Transactions on Sensor Networks (TOSN),* vol. 4, p. 15, 2008.
[14] Y. Sun, Z. Han, and K. R. Liu, "Defense of trust management vulnerabilities in distributed networks," *Communications Magazine, IEEE,* vol. 46, pp. 112-119, 2008.
[15] Y. Wang, R. Chen, and J.-H. Cho, "Trust-based Service Management of Mobile Devices in Ad Hoc Networks."
[16] S. Buchegger and J.-Y. Le Boudec, "A robust reputation system for mobile ad-hoc networks," 2003.
[17] K. Kane and J. C. Browne, "Using uncertainty in reputation methods to enforce cooperation in ad-hoc networks," presented at the Proceedings of the 5th ACM workshop on Wireless security, Los Angeles, California, 2006.
[18] Y. N. Liu, K. Q. Li, Y. W. Jin, Y. Zhang, and W. Y. Qu, "A novel reputation computation model based on subjective logic for mobile ad hoc networks," *Future Generation Computer Systems-the International Journal of Grid Computing and Escience,* vol. 27, pp. 547-554, May 2011.
[19] H. Lin, J. Ma, J. Hu, and K. Yang, "PA-SHWMP: a privacy-aware secure hybrid wireless mesh protocol for IEEE 802.11 s wireless mesh networks," *EURASIP Journal on Wireless Communications and Networking,* vol. 2012, pp. 1-16, 2012.
[20] R. Chen, J. Guo, F. Bao, and J.-H. Cho, "Trust management in mobile ad hoc networks for bias minimization and application performance maximization," *Ad Hoc Networks,* vol. 19, pp. 59-74, 2014.

[21]   L. Ruidong, L. Jie, and H. Asaeda, "A Hybrid Trust Management Framework for Wireless Sensor and Actuator Networks in Cyber-Physical Systems," *IEICE TRANSACTIONS on Information and Systems,* vol. 97, pp. 2586-2596, 2014.

[22]   Y. Wang, Y.-C. Lu, I.-R. Chen, J.-H. Cho, A. Swami, and C.-T. Lu, "LogitTrust: A logit regression-based trust model for mobile ad hoc networks," 2015.

[23]   G. Shafer, *A Mathematical Theory of Evidence*: Princeton University Press, 1976.

[24]   C. Perkins, E. Royer, and S. Das, "RFC 3561 Ad hoc On-Demand Distance Vector (AODV) Routing," 2003.

[25]   A. Jsang and R. Ismail, "The beta reputation system," in *Proceedings of the 15th bled electronic commerce conference*, 2002, pp. 2502-2511.

[26]   A. Jøsang, "Artificial reasoning with subjective logic," in *Proceedings of the Second Australian Workshop on Commonsense Reasoning*, 1997.

[27]   A. Josang, E. Gray, and M. Kinateder, "Simplification and analysis of transitive trust networks," *Web Intelli. and Agent Sys.,* vol. 4, pp. 139-161, 2006.

[28]   A. Jøsang, R. Hayward, and S. Pope, "Trust network analysis with subjective logic," in *Proceedings of the 29th Australasian Computer Science Conference-Volume 48*, 2006, pp. 85-94.

[29]   S. Dietzel, R. van der Heijden, H. Decke, and F. Kargl, "A flexible, subjective logic-based framework for misbehavior detection in V2V networks," in *World of Wireless, Mobile and Multimedia Networks (WoWMoM), 2014 IEEE 15th International Symposium on a*, 2014, pp. 1-6.

[30]   N. Ben Salem and J. P. Hubaux, "Securing wireless mesh networks," *Wireless Communications, IEEE,* vol. 13, pp. 50-55, 2006.

[31]   *NS-3 network simulator homepage*. Available: http://www.nsnam.org